\documentclass[11pt]{article}

\PassOptionsToPackage{numbers, compress}{natbib}

\usepackage[preprint]{neurips_2019}

\usepackage[utf8]{inputenc} 
\usepackage[T1]{fontenc}    
\usepackage{hyperref}       
\usepackage{url}            
\usepackage{booktabs}       
\usepackage{amsfonts}       
\usepackage{nicefrac}       
\usepackage{microtype}      
\usepackage{graphicx}
\usepackage{float}

\title{Pattern and Anomaly Detection in Urban Temporal Networks}

\author{%
\mbox{
 Mingyi He$^1$\thanks{Mingyi He,Shivam Pathak and Urwa Muaz contributed equally to the paper,the list is alphabetically ordered.}~~~
  Shivam Pathak$^1$\footnotemark[1]~~~
  Urwa Muaz$^1$\footnotemark[1]~~~}\\
  \mbox{
  \bf Jingtian Zhou$^1$~
  Saloni Saini$^1$~
  Sergey Malinchik$^2$~
  Stanislav Sobolevsky$^1$~}\\
  $^1$Center for Urban Science and Progress\\
  New York University\\
  New York, NY, 11201 \\
  $^2$Lockheed Martin Advanced Technology Lab\\
  Lockheed Martin Corporation\\
  Cherry Hill, NJ, 08002
}

\begin{document}

\maketitle
\begin{flushleft}
\begin{abstract}
Broad spectrum of urban activities including mobility can be modeled as temporal networks evolving over time. Abrupt changes in urban dynamics caused by events such as disruption of civic operations, mass crowd gatherings, holidays and natural disasters are potentially reflected in these temporal mobility networks. Identification and early detecting of such abnormal developments is of critical importance for transportation planning and security. 
Anomaly detection from high dimensional network data is a challenging task as edge level measurements often have low values and high variance resulting in high noise-to-signal ratio. In this study, we propose a generic three-phase pipeline approach to tackle curse of dimensionality and noisiness of the original data. Our pipeline consists of i) initial network aggregation leveraging community detection ii) unsupervised dimensionality reduction iii) clustering of the resulting representations for outlier detection. We perform extensive experiments to evaluate the proposed approach on mobility data collected from two major cities, New York City and Taipei. 
Our results empirically prove that proposed methodology outperforms traditional approaches for anomaly detection.
We further argue that the proposed anomaly detection framework is potentially generalizable to various other types of temporal networks e.g. social interactions, information propagation and  epidemic spread.
\end{abstract}


\section{INTRODUCTION}
Anomaly detection can be simply defined as the identification of observations that deviate significantly from the normal patterns\cite{zimek2017outlier}. In urban systems, besides mitigating operational disruptions it has applications in the domain of public safety for early detection of threats and suspicious activities. Examples of these scenarios includes 2015 New Year’s Eve celebration in Shanghai, where overcrowding resulted in a stampede causing 36 casualties. Similarly, in 2016 Central Park, New York City witnessed a dangerous stampede of ‘Pokemon Go’ players who gathered in hundreds to pursue a game character. Early detection of these events can enable the authorities to take preventive measures to effectively mitigate the adverse consequences. In various research works \cite{weinberger2011spies,bohannon2012tweeting, abdelhaq2013eventweet, chen2014non}  social media data has been utilized for the detection of civic unrest. Bohannon discusses that social media data can be leveraged to forecast civic riots\cite{bohannon2012tweeting}. Abdelhaq et al \cite{abdelhaq2013eventweet} proposes a novel framework for localized event detection from twitter data. In \cite{abdelhaq2013eventweet, chen2014non}, scan statistics was used on social media data for event detection. However, many urban events have an effect which is localized in space, necessitating the need to use geospatial urban data such as human mobility. An urban phenomenon like human mobility can be represented using temporal networks, studying these networks can provide valuable insights about normal urban behaviors and can potentially help to detect anomalous events. 
Temporal networks are dynamic in nature and are constantly subject to change with time, the temporal change could be in attributes alone, in structure alone or both. In this study, we limit ourselves to networks for which only edge values evolve over time. 
Network anomalies can be of different types and scales, and can be limited to a subset of nodes or edges, but in this study we are interesting in identifying the points in time when the whole network state can be regarded as anomalous. This type of anomalies are called event anomalies.

In the urban context, the number of anomalous days are always significantly smaller in comparison to that of normal days. Due to this scarcity of anomalies, supervised methods will be prone to over-fitting and will generalize poorly to new types of anomalies\cite{bellman1966dynamic}. Unsupervised event detection can be performed by assigning a likelihood score to network states at each point of time and then classifying states with extremely low scores as anomalies\cite{hodge2004survey}\cite{laxhammar2008anomaly}. Application of probabilistic methods directly on all nodes and edges of the network is not fruitful as information at such granularity has a very high noise to signal ratio, inhibiting the detection of useful patterns\cite{kriegel2009clustering}. A legacy approach for this is a system-wide aggregation of ridership, followed by time series models to calculate residuals, and then detecting anomalies using the probabilistic method\cite{guralnik1999event}. Although system-wide aggregation helps to address the fluctuation in the whole network at once, relevant signals are lost and it fails to account for changes in local patterns of the network. A slightly more sophisticated approach uses node wise aggregation of incoming and outgoing ridership. This is able to accommodate local patterns changes but fails to address changes in structural patterns. 

To mitigate the limitations of these approaches, we aim to develop a method that appropriately captures local phenomena at different spatial scale and the structural patterns emerging in the networked data. To achieve this objective, we build upon the method proposed by Stanislav et al. (2019)\cite{Stan19} for anomaly detection. The method has three main phases: (1) Community Detection for coarsening the network, (2) Dimensionality reduction to learn a low-dimensional representation, and (3) Detect anomalies using the learned representation.

Applied potential of such an approach in transportation and beyond is magnified by the abundance of urban data reflecting various aspects of human activity. Cell phone \cite{Sobolevsky2013delineating,kung2014exploring,amini2014impact,pei2014new,grauwin2017identifying} and landline connections \cite{Ratti2010GB}, credit card transactions \cite{sobolevsky2016cities,hashemian2017socioeconomic}, GPS readings \cite{santi2014quantifying}, geo-tagged Twitter \cite{hawelka2014geo,belyi2017global} and Flickr data \cite{paldino2015urban}, 311 service requests \cite{wang2017structure} and Bluetooth connections \cite{yoshimura2014analysis} have all been demonstrated to be useful proxies for urban dynamics revealing different aspects of it \cite{zhu2018digital}. In the present paper we will evaluate the proposed approach on taxi trip data for New York City and subway travel data for Taipei. 

\section{RELATED METHODS}
There is an abundance of related work revolving around the individual components of our pipeline approach described above, but cascading these approaches has not been widely studied. In this section we discuss the methods that deal with event detection in temporal networks and are closely related to individual components in our pipeline. We explain how our methods builds upon these techniques and where they differ from them.

\subsection{Probabilistic Clustering Models}
These are techniques that construct a probabilistic model of normal behavior and identify anomalies as observations with small likelihood under the model. For a simple example could one could fit a multivariate probability distribution like Gaussian to the data and perform outlier detection based on p-value threshold \cite{hodge2004survey}. But real world data often have multiple underlying distributions and assumption of a single distribution does not hold. In our case, e.g. the weekdays are expected to have a different ridership distribution than weekends. A potential solution is to use Gaussian mixture models (GMM) \cite{reynolds2015gaussian}, which is parametric probability distribution model which represents the data distribution as weighted sum of normally distributed sub-populations. Outlier detection can then be performed by p-value thresholding on component sub-populations \cite{laxhammar2008anomaly}. We decided to use GMMs for anomaly detection, but we note that it is not a good idea to apply it directly to our high dimensional network data. GMMs do not perform well with high dimensional data, especially when the number of observations is not many times larger than the number of dimensions \cite{kriegel2009clustering}. This is a generic issue with machine learning models and is referred to as curse of dimensionality \cite{bellman1966dynamic}. Using a subset of most useful features or applying dimensionality reduction techniques prior to GMMs are possible solutions to this problem \cite{kriegel2009clustering}. So we use GMM as the last stage in our pipeline, performing topological aggregation (community detection) and dimension reduction before it to reduce dimension. These techniques are introduced below.

\subsection{Community Detection}
Community detection is a technique for topological aggregation of the networks, which could be also used in anomaly detection since community partitioning observed over time can capture strong structural changes. In our approach we use community detection for node aggregation to reduce noise to signal ratio which is high at individual edge level measurements. In simple terms, communities in graphs can be described as sets of vertices that share common properties and play similar roles in the graphs \cite{fortunato2010community}. There is usually a high concentration of edges within the community as compared to edge concentration between communities. An alternative approach to community detection for grouping similar nodes is clustering using algorithms like k-means \cite{macqueen1967some} and DBSCAN \cite{ester1996density}. Clustering algorithms would normally group the nodes only on the base of their values, but network has information about connectivity of the nodes and connectivity also often correlates with spatial proximity. Community detection leverages all these properties to group similar nodes unlike clustering, so it is preferable in our case. A new algorithm COMBO recently proposed in \cite{sobolevsky2014general} presents a strategy to search the possible partition space with objective of optimizing a specific objective function, modularity being the most common one. COMBO outperforms other state-of-the-art approaches in most cases while maintaining reasonable execution times. Hence, we use COMBO for community detection in this study. Since the community detection captures the topological characteristics of the network, the partition of the network achieved by community detection can provide insight into the state of network. A metric based on the similarity of the partition among network states can be used as vector representation that will be used for anomaly detection. For example, sudden change in the number of communities or membership of communities can be identified as an anomalous event. \cite{gupta2012integrating} analyzes change in community belongingness of the vertices over time and identifies the vertices whose belongingness change is different from average as anomalous. \cite{duan2009community}  analyzes partitioning over time and uses a similarity metric based on Jackard coefficient between respective partitions. Subsequently, they perform a threshold based event detection on this metric. In contrast to these methods, we use community detection as the initial step in the pipeline to reduce the dimensionality of the data and mitigate noise while preserving topological information that can be useful for anomaly detection. 

\subsection{Decomposition dimensionality reduction}

Decomposition is a an approach to perform dimensionality reduction to facilitate the use of traditional anomaly detection techniques and avoid curse of dimenstionality described above. This technique leverages the concept of tensor decomposition which can be used to represent multidimensional matrices in lower dimensions. Similar to compression techniques, regularities in data are utilized to achieve this. The initial step of this approach is to represent the network as a matrix, which can be done by adding a dimension for every measure of interest. In terms of temporal transport network, dimensions could be origin, destination and time. Most widely used means of achieving matrix decomposition are PCA \cite{jolliffe2011principal} and SVD \cite{golub1971singular}. These are linear transformations for which an inverse transformation also exists, hence reconstruction from reduced space can be performed and it will have an error associated with it.  Decomposition can be used in two ways: 1) reconstruction error is analyzed over time and the time points where error is above a threshold are considered anomalous\cite{miller2012scalable}. Rationale here is that anomalous behavior would inhibit decomposition. 2) the reduced representation is analyzed over time to discover anomalies\cite{lakhina2004diagnosing}, assuming that anomalies would show up in reduced space. We note that both SVD and PCA are limited to linear decomposition, and also explore the space of nonlinear decomposition techniques using deep autoencoders for representation learning\cite{hinton2006reducing}. The representation space will then be used to detect outliers using GMM described above.

\section{DATA AND PREPROCESSING}
 Evaluating the performance of the anomaly detection approaches could be seen as challenge in the absence of the ground truth labeled anomalies. Our evaluation is based on a fundamental assumption that certain major events including holidays or weather events should change the daily transportation patterns significantly. We regard these events as anomalous and evaluate the tested methodologies on their ability to identify these events as anomalies in space of transportation data. Hence, for the detection of anomalies in urban mobility network, we have used two categories of data sets - Urban mobility data sets and Event data sets.

\subsection{Urban Mobility Data Sets}
The urban mobility data sets were collected for the cities of interest, including taxi ridership for New York (USA) and subway ridership for Taipei (China). We have aggregated all these data sets at the daily level and have transformed them into a convenient uniform format which is described in Table \ref{format}. The summary of the aggregated mobility data sets is presented in Table \ref{mobility}.

\begin{table}[!htbp]
\centering
\caption{Summary of aggregated mobility data sets.}
\begin{tabular}{@{}ccccccc@{}}
\toprule
\textbf{\begin{tabular}[c]{@{}c@{}}Data \\ Source\end{tabular}} & \textbf{Timeframe} & \textbf{\begin{tabular}[c]{@{}c@{}}Temporal \\ Points\end{tabular}} & \textbf{\begin{tabular}[c]{@{}c@{}}Ridership \\ Collected\end{tabular}} & \textbf{\begin{tabular}[c]{@{}c@{}}Avg Ridership \\ (per station\\ per day)\end{tabular}} & \textbf{\begin{tabular}[c]{@{}c@{}} No. of Nodes\\ (Stations/ \\ Zones)\end{tabular}} & \textbf{\begin{tabular}[c]{@{}c@{}}No. of \\ Edges\end{tabular}} \\ \midrule
Taipei Subway & \begin{tabular}[c]{@{}c@{}}2017-01-01\\ to \\ 2018-12-31\end{tabular}&637 & 1307013573 & 18969 & 108 & 11664 \\
New York Taxi & \begin{tabular}[c]{@{}c@{}}2017-06-01\\ to\\ 2018-12-31\end{tabular}& 580 & 481656030 & 2815 & 263 & 65792 \\ \bottomrule
\end{tabular}
\label{mobility}
\end{table}

\subsection{Events Data Sets}
To benchmark the efficacy of our method in detecting events where the legacy methods perform well, and to also detect events where they fail, we have selected a set of global and significant local events for this study. The different types of events we have considered are National Holidays, Cultural Events, Parades, Protests, and Extreme Weather. The details and summary statistics for these data sets are present in Table \ref{events}. The description for this is present in Table \ref{format}. 

\begin{table}[H]
\centering
\caption{Summary of aggregated events data sets.}
\begin{tabular}{@{}cccc@{}}
\toprule
\textbf{Cities} & \textbf{Natural Disaster} & \textbf{National Holiday} & \textbf{Cultural Event} \\ \midrule
Taipei & 3 & 26 & 4  \\
New York City & 3 & 16 & 3  \\ \bottomrule
\end{tabular}
\label{events}
\end{table}

\subsection{Details of Data Format}

This section provides the details about the sources of data sets used in the study, and the preprocessing performed to create extreme weather events. Table \ref{format} describes the standard format of data used in this study.

\begin{table}[H]
\centering
\caption{Standard data formats used in this project}
\begin{tabular}{@{}ccc@{}}
\toprule
\textbf{Type of Data} & \textbf{Data sets} & \textbf{Data Format} \\ \midrule
Urban Mobility Data & Taxi / Subway Ridership Data set & \begin{tabular}[c]{@{}c@{}}Date, Start\_id, End\_id, \\ Volume of Flow\end{tabular} \\
Event Data & \begin{tabular}[c]{@{}c@{}}National Holiday, Cultural Events, \\ and Natural Disasters \end{tabular} & Date, Type of Event, Description \\ \bottomrule
\end{tabular}
\label{format}
\end{table}

\subsection{Description for Culture Event}
Culture events include sports events, protests, parade, concert, etc, and usually regarded as local events which means can only be detected as anomaly in certain area. We collected data from the list of biggest protests in United States \cite{wiki:protest}, and annually sports events in both Taipei and New York, and annually concert in Taipei. It is impossible to collect all the local anomalies in a city, but we believe the days we selected are typical culture events.
\begin{table}[H]
\centering
\caption{Collected Culture Events Days}
\begin{tabular}{@{}cccc@{}}
\toprule
\textbf{City} & \textbf{Date} & \textbf{Culture Event Type} & \textbf{Description} \\ \midrule
New York & 2017-09-10 & Sports Event & Most viewership day in 2017 U.S Open \\
New York & 2018-01-20 & Protest & Women's March, more than 1,500,000 participants \\
New York & 2018-09-08 & Sports Event & Most viewership day in 2017 U.S Open \\
Taipei & 2017-06-24 & Concert & The 28th Golden Melody Awards \\
Taipei & 2017-11-25 & Concert & The 54th Golden Horse Awards \\
Taipei & 2017-12-17 & Sports Event & Marathon \\
Taipei & 2018-06-23 & Concert & The 29th Golden Melody Awards \\ \bottomrule
\end{tabular}
\end{table}
\subsection{Timeseries Visualization of Data}
Figure \ref{timeseries} provides a simple visualization of aggregated daily ridership and collected events data for the city of Taipei. It can be seen that a good fraction of the events tend to lie on extreme ends of the ridership.
\begin{figure}[!htbp]
\centering
\caption{Taipei aggregated time series}
\includegraphics[width=13cm]{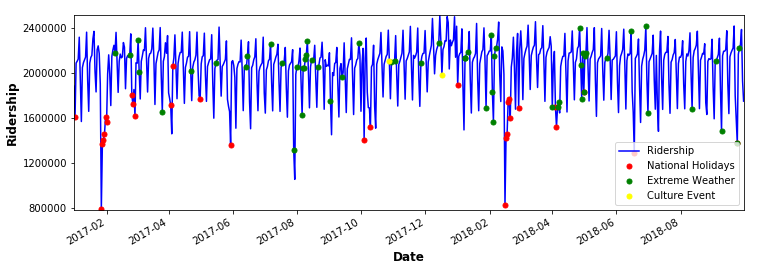}
\label{timeseries}
\end{figure}

\section{METHODOLOGY}
We represent the urban mobility data as a temporal network where nodes represent places/stations, and edges correspond to the flow of people between the nodes. Here, only the edge value changes based on the ridership at any given time. Throughout this study, we refer point: $P(N,E,t)$, as the network corresponding to any given time $t$ with $N$ nodes and $E(N * N)$ edges. Building upon the method proposed in Stanislav et al. (2019)\cite{Stan19}, to detect event anomalies amongst these points we assign likelihood score to each using the probabilistic method/Gaussian Mixture Model.  To use GMM, we first coarsen network using community detection followed by further decomposition dimensionality reduction.

\subsection{Probabilistic Outlier Detection using GMM}

To calculate the probabilistic score of each point to be classified as an outlier, we use Gaussian mixture models(GMM) and cluster the dominant patterns of ridership. Next, with this GMM, we estimate the likelihood of each point to be in one of the defined clusters. Points with likelihood less than a certain threshold are classified as anomalous. These anomalous points are removed, and the clustering is rerun on the reduced set of points. Using this new clustering model, we re-estimate the likelihood, detect anomalies, and iteratively repeat this whole process until the convergence is reached and number of anomalous points does not increase  with iterations.

GMM does not perform well with high dimensional data, especially when the items to attributes ratio is significantly low\cite{bouveyron2007high}. Through our experiment(Figure \ref{GMM}), we established that increasing the number of dimension above 200 in the input of GMM leads to an exponential increase in run-time, and saturation in performance improvement. This inhibits us from directly using GMM on high dimensional network data and necessitates the need for aggregation. Network aggregation reduces the dimension and enhances the signal to noise ratio, this can be done using the overall aggregation of ridership, node wise aggregation of incoming and outgoing ridership, or community detection based aggregation of nodes.

\begin{figure}[!htbp]
\centering
\caption{GMM Input Dimensions vs Run Time and Performance: The plot shows the relation between input dimensions of Gaussian Mixture Model(GMM), and its run time and performance.  As we increase the dimensions in the input of the GMM, the run time increases exponentially, and the F1 performance for detecting anomalies starts decreasing after a certain limit. This inhibits us from using GMM directly on high dimensional network data, and is also known as curse of dimensionality.}
\includegraphics[width=13cm]{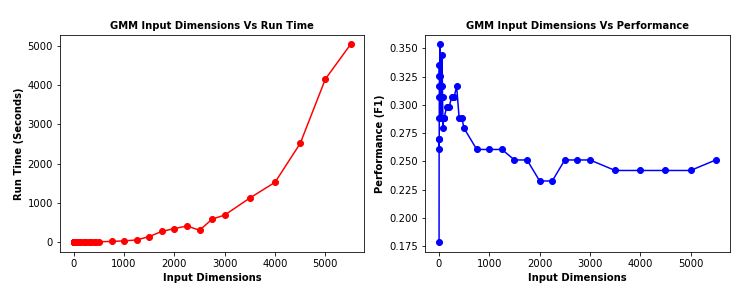}
\label{GMM}
\end{figure}

\subsection{Network Aggregation}
The two main problems in urban mobility networks are data sparsity, and low signal to noise ratio. The data sparsity is caused by the presence of remote stations/places and limited riders accessing them. As these networks are huge in size and have a large number of edges, low signal to noise ratio at edge level measurements makes anomaly detection an intractable task. Thus, it is necessary to coarse this network to reduce its dimensions. Network could be aggregated by overall aggregation, node-wise aggregation, and community detection.

\textbf{Node-wise Aggregation:} In node-wise aggregation, we sum the incoming and outgoing ridership of each node thus, reducing the dimensions from $N * N$ to $2N$ (N: Number of nodes in the network). Real-world mobility networks usually have 100+ nodes, making 2N still large enough to be used as an input for GMM.  Thus, we apply dimension reduction techniques such as PCA on this reduced network before applying the GMM for anomaly detection.

\textbf{Community Detection:} Rather than using crude methods of aggregation, the incumbent method used in literature to reduce the size of urban transportation network is that of community detection. In community detection, for each edge in the original network, the algorithm compares the actual edge weight with the average expected value. The edges with positive relative strength scores represent particularly strong network connections and are placed inside the community, while edges with negative score are placed between the communities. This process is done by maximizing the modularity score and determining the optimal partitioning.
 In a directed network graph for each edge $e(x,y)$ in the set of edges E; the relative strength score $q(x,y)$is calculated using the formula:
 
\begin{equation}
    q(x,y) = \frac{e(x,y)}{T} - \frac{k^{x}_{out}k^{x}_{in}}{T^{2}} 
\end{equation}

Where, \newline
$q(x,y)$	: The relative strength score between two nodes x and y,\newline
$e(x,y)$: The  edge weight/ridership between two nodes x and y, \newline
$T$: Total network weight/ridership, \newline
$k^{x}_{out}= \sum_{i=1}e(x,i)$: Outgoing strength of the node $x$, \newline
$k^{x}_{in}= \sum_{i=1}e(i,y)$: Incoming strength of the node $y$. \newline

Then given a certain partitioning $P=(C_x,x \in N)$, a vector of discrete community numbers  $C_x$  associated with each node $x$ of the set $N$(Original nodes in the network), one can define the overall modularity score:

\begin{equation}
    Q(P) = \sum_{x,y,cx=cy} q(x,y) = \sum_{x,y, cx=cy}[\frac{e(x,y)}{T} - \frac{k^{x}_{out}k^{x}_{in}}{T^{2}}]
\end{equation} 
A trivial partitioning $P_0$ considers the entire network as a single community and results in modularity score $Q(P) = 0$. Thus any reasonable partitioning better than a trivial one will have the modularity score greater than 0 with an upper limit of 1. We solve this problem of increasing modularity score and optimizing the partition P using the optimization technique COMBO introduced by Sobolevsky et al[10]. The community detection algorithm reduces the size of the network from $N * N$ to $C * C$, (where $N$ is the number of nodes in the original network, and $C$ is the number of communities).

The communities can further be partitioned into sub communities by applying community detection on each community. We use this nested community detection approach to achieve communities of desired spatial granularity.

\subsection{Dimensionality Reduction}

The size of the original network is considerably reduced by the network aggregation. Although, many people directly use this reduced network for anomaly detection. We found that the reduced networks are still very high dimensional and this can inhibit effective application of probabilistic models for anomaly detection. Thus, in our pipeline we use decomposition techniques to learn a compact representation of the network further reducing the dimensionality. We conduct the linear dimension reduction using principal component analysis, and non-linear dimension reduction using deep autoencoder. Principal component analysis (PCA) is a linear decomposition, which is based on the eigen‐decomposition of positive semi‐definite matrices and on the singular value decomposition (SVD) of rectangular matrices, to extract the important information from data sets\cite{abdi2010principal}. Autoencoders are neural networks with two components: encoder and decoder. It compresses the information in original mobility network to a reduced dimension space using the encoder and decompresses the data back to original dimensions with the help of the decoder. A well trained autoencoder yields a nonlinear compressed representation of the original urban mobility network.   

\subsection{Baseline}
 Evaluation of the newly proposed approach requires a baseline. A first one is based on overall aggregation of ridership and time series modelling. For overall aggregation, we sum the total ridership in the network from each point in each day. The time series for aggregated ridership of Taipei Subway Network is shown in figure \ref{timeseries}. Further, this time series is modeled using ARIMA model with periodicity of 7 days, and the residuals is obtained by subtracting the predicted ridership from the original. Thirdly, we applied the GMM on the residuals to detect anomalies.
 
 Another baseline approach uses a common network aggregation into inflow/outflow (IO) per each node. We will use IO node wise aggregation followed by dimensionality reduction and GMM as a baseline for evaluating comparative utility of the initial network topological aggregation through community detection.

\section{METHOD COMPARISON}
To benchmark the performance of our proposed methodology on real-world urban anomalies detection, we design experiments using our datasets. We compare the results of our three stage pipeline to two staged approach where we omit the dimensionality reduction step. This is designed to shed light on the need for further dimensionality reduction after network aggregation. Additionally, we also compare the results to the time series baseline described above. Furthermore, we also experiment with two types of aggregation techniques and two type of dimensionality reduction methods within the framework of our pipeline to find the optimal cascade of methods for our data. Summary of the 7 tested methods is provided in Table \ref{method}.

\begin{table}[H]
\caption{Methods Description}
\centering
\begin{tabular}{lll}
\toprule
\textbf{Method Name} & \textbf{Description} \\ \midrule
1.1: Time Series & Overall Aggregation+ARIMA Residual+GMM \\
2.1: IO & Node wise Inflow-Outflow Aggregation + GMM \\
2.2: IO + PCA & Node wise Inflow-Outflow Aggregation + PCA + GMM \\
2.3: IO + AE & Node wise Inflow-Outflow Aggregation + Autoencoders + GMM \\
3.1: Comm & Nested Community Detection + GMM \\
3.2: Comm + PCA & Nested Community Detection + PCA + GMM \\
3.3: Comm + AE & Nested Community Detection + Autoencoders + GMM \\ \bottomrule
\end{tabular}
\label{method}
\end{table}

\subsection{Hyperparameter Selection}

For GMM Component Selection we run GMM with 1 to 5 components on network data and use Bayesian Information Criterion (BIC) \cite{scikit-learn} criteria for selection of optimal components. As the dimension of raw network data is too high (see table \ref{mobility}) to run GMM directly on it, we applied PCA to reduce the dimension to 15 before tuning the number of components. We use representation space of 15 for PCA which was designed to account for about 80 percent of the variance of data. We use similar dimension of representation space for autoencoder for fair comparison. Autoencoder hyperparameters were selected by using random search over the predefined space of hyperparameters and optimizing for reconstruction cost. In our current autoencoder architecture, for both encoder and decoder we use three hidden layers with 75, 25 \& 15 neurons respectively in each. Adams optimizer with cross entropy loss is used for training.

\subsection{Evaluation Metrics}
We used Area Under the Receiver Operating Characteristic curve (AUROC) to evaluate the algorithms' performances. AUROC measures the classification performance under different threshold, the higher AUROC value means the better results in terms of true positive rate under comparable false positive rate. Usually, for a binary classification, the AUROC value for random choice is 0.5. This is a suitable metric because it is agnostic of the anomaly threshold and gives aggregated performance over the entire range of thresholds.

\section{RESULTS AND DISCUSSION}

The performance of all these methods on the two considered cities is reported on Figure \ref{ROC} and Table \ref{AUC}. By comparison of AUROC the proposed pipeline, Comm + PCA, has the highest AUROC value in both Taipei and New York. High AUROC values for both cities highlights applicability of the pipeline approach to various data sets and geographies. Additionally, all the pipeline configurations yield superior performance to the time series baseline. This is because the pipeline approach preserves the topological information in the data while baseline uses crude aggregation which discards this information. Pipeline approach is equipped to detect localized anomalies which break the correlation patterns of data set without affecting the overall ridership while baseline approach lacks in this aspect. Furthermore, adding the third stage of dimensionality reduction over the network aggregation enhances the performance in all experiments irrespective of the type of network aggregation. This is because of inability of GMMs to address the problem of high dimensionality in aggregated networks. This highlights the importance of three staged pipeline approach proposed by this study as opposed to a more common two staged approach.

The results establish that using community detection as the aggregation step results in superior performance to node wise aggregation. This is because community detection leverages rich information from origin-destination network data to perform graph aggregation that preserves topological and structural information. Node level aggregation is able to accommodate local patterns changes but fails to address dynamics of structural patterns. It is interesting to note that non-linear dimensionality reduction using autoencoder gave inferior performance to PCA for most experiments, except of IO+AE. This could be due to the absence of non-linearity in the data or difficulty of training a reliable autoencoder for this modest sized data set.

\begin{figure}[H]
\centering
\caption{Comparison of the AUROC curve from different methods in different cities.}
\includegraphics[width=14cm]{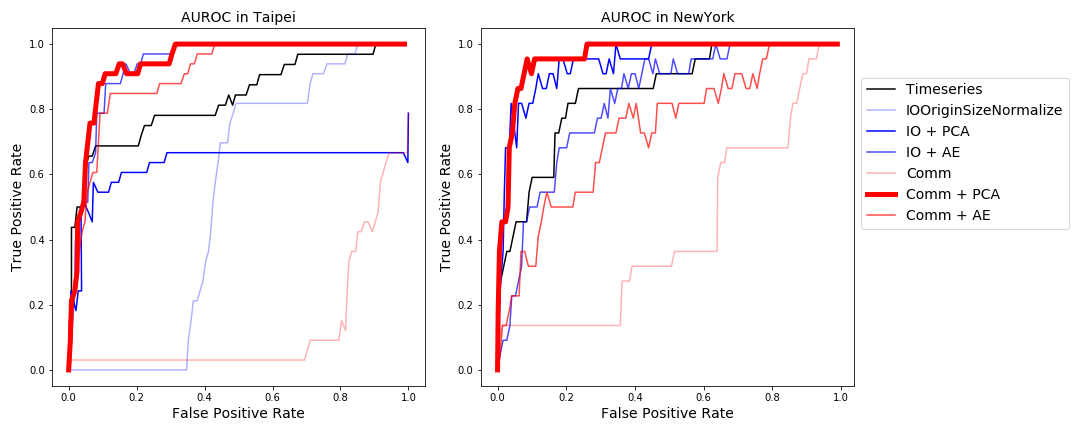}
\label{ROC}
\end{figure}

\begin{table}[!htbp]
\caption{AUC value from different methods in two cities}
\centering
\begin{tabular}{ccc}
\hline
\textbf{Method} & \textbf{AUROC Taipei} & \textbf{AUROC New York} \\ \hline
Time Series & 0.53 & 0.70 \\
IO & 0.51 & 0.56 \\
IO + PCA & 0.63 & 0.93 \\
IO + AE & 0.92 & 0.81 \\
Comm & 0.13 & 0.41 \\
\textbf{Comm + PCA} & \textbf{0.93} & \textbf{0.95} \\
Comm + AE & 0.90 & 0.72 \\ \hline
\end{tabular}
\label{AUC}
\end{table}

\section{CONCLUSION}

We evaluated several methods for anomaly detection over temporal networks of human mobility, including a proposed three-phase pipeline approach combining initial network aggregation leveraging community detection, followed by unsupervised dimensionality reduction and Gaussian Mixture clustering of the resulting network representations for outlier detection. Performance of the anomaly detection has been tested against the ground truth labeled samples of known events. It is however important to understand that these samples are not exhaustive and some of the false positives might still be valid anomalies.

Experiments on real-world taxi data in New York and subway data in Taipei exhibited that the proposed pipeline approach by preserving topological structure and connectivity of networks, outperforms both baseline methods based on the overall mobility aggregation and node-wise spatial aggregation. Both, initial network aggregation and further dimensionality reduction appear to be important steps as final Gaussian Mixture Model has a better performance on low-dimensional data sets. The baseline time series analysis of daily aggregations does well in isolating anomalies which have a clear global impact while it fails in isolating localized anomalies.  Experiments on real-world data have not revealed consistent advantage of Autoencoder over PCA for the dimensionality reduction step, with certain exceptions like handling node-based network aggregation for Taipei. In general, while proposed pipeline approach gained an overall numerical advantage in our evaluation experiments, specific cases often have different methods demonstrating the best performance. Further evaluation on larger samples of labeled real-world event in diverse geographies and data sources may provide further details in diagnostics of comparative capabilities of different methodologies in isolating different kinds of anomalies.

\bibliographystyle{unsrt}
\end{flushleft}
\bibliography{ref.bib}

\begin{thebibliography}{10}

\bibitem{zimek2017outlier}
Arthur Zimek and Erich Schubert.
\newblock Outlier detection.
\newblock {\em Encyclopedia of Database Systems}, pages 1--5, 2017.

\bibitem{weinberger2011spies}
Sharon Weinberger.
\newblock Spies to use twitter as crystal ball, 2011.

\bibitem{bohannon2012tweeting}
John Bohannon.
\newblock Tweeting the london riots, 2012.

\bibitem{abdelhaq2013eventweet}
Hamed Abdelhaq, Christian Sengstock, and Michael Gertz.
\newblock Eventweet: Online localized event detection from twitter.
\newblock {\em Proceedings of the VLDB Endowment}, 6(12):1326--1329, 2013.

\bibitem{chen2014non}
Feng Chen and Daniel~B Neill.
\newblock Non-parametric scan statistics for event detection and forecasting in
  heterogeneous social media graphs.
\newblock In {\em Proceedings of the 20th ACM SIGKDD international conference
  on Knowledge discovery and data mining}, pages 1166--1175. ACM, 2014.

\bibitem{bellman1966dynamic}
Richard Bellman.
\newblock Dynamic programming.
\newblock {\em Science}, 153(3731):34--37, 1966.

\bibitem{hodge2004survey}
Victoria Hodge and Jim Austin.
\newblock A survey of outlier detection methodologies.
\newblock {\em Artificial intelligence review}, 22(2):85--126, 2004.

\bibitem{laxhammar2008anomaly}
Rikard Laxhammar.
\newblock Anomaly detection for sea surveillance.
\newblock In {\em 2008 11th international conference on information fusion},
  pages 1--8. IEEE, 2008.

\bibitem{kriegel2009clustering}
Hans-Peter Kriegel, Peer Kr{\"o}ger, and Arthur Zimek.
\newblock Clustering high-dimensional data: A survey on subspace clustering,
  pattern-based clustering, and correlation clustering.
\newblock {\em ACM Transactions on Knowledge Discovery from Data (TKDD)},
  3(1):1, 2009.

\bibitem{guralnik1999event}
Valery Guralnik and Jaideep Srivastava.
\newblock Event detection from time series data.
\newblock In {\em Proceedings of the fifth ACM SIGKDD international conference
  on Knowledge discovery and data mining}, pages 33--42. ACM, 1999.

\bibitem{Stan19}
Stanislav Sobolevsky, Philipp Kats, Colin Bradly, Mingyi He, and Sergey
  Malinchik.
\newblock Anomaly detection in temporal networks.
\newblock Accepted for NetSci 19, 1 2018.

\bibitem{Sobolevsky2013delineating}
Stanislav Sobolevsky, Michael Szell, Riccardo Campari, Thomas Couronn{\'e},
  Zbigniew Smoreda, and Carlo Ratti.
\newblock Delineating geographical regions with networks of human interactions
  in an extensive set of countries.
\newblock {\em PloS one}, 8(12):e81707, 2013.

\bibitem{kung2014exploring}
Kevin~S Kung, Kael Greco, Stanislav Sobolevsky, and Carlo Ratti.
\newblock Exploring universal patterns in human home-work commuting from mobile
  phone data.
\newblock {\em PloS one}, 9(6):e96180, 2014.

\bibitem{amini2014impact}
Alexander Amini, Kevin Kung, Chaogui Kang, Stanislav Sobolevsky, and Carlo
  Ratti.
\newblock The impact of social segregation on human mobility in developing and
  industrialized regions.
\newblock {\em EPJ Data Science}, 3(1):6, 2014.

\bibitem{grauwin2017identifying}
Sebastian Grauwin, Michael Szell, Stanislav Sobolevsky, Philipp H{\"o}vel,
  Filippo Simini, Maarten Vanhoof, Zbigniew Smoreda, Albert-L{\'a}szl{\'o}
  Barab{\'a}si, and Carlo Ratti.
\newblock Identifying and modeling the structural discontinuities of human
  interactions.
\newblock {\em Scientific reports}, 7:46677, 2017.

\bibitem{Ratti2010GB}
Carlo Ratti, Stanislav Sobolevsky, Francesco Calabrese, Clio Andris, Jonathan
  Reades, Mauro Martino, Rob Claxton, and Steven~H Strogatz.
\newblock {Redrawing the Map of Great Britain from a Network of Human
  Interactions}.
\newblock {\em PLoS ONE}, 5(12):e14248, 2010.

\bibitem{sobolevsky2016cities}
Stanislav Sobolevsky, Izabela Sitko, Remi~Tachet Des~Combes, Bartosz Hawelka,
  Juan~Murillo Arias, and Carlo Ratti.
\newblock Cities through the prism of people’s spending behavior.
\newblock {\em PloS one}, 11(2):e0146291, 2016.

\bibitem{hashemian2017socioeconomic}
Behrooz Hashemian, Emanuele Massaro, Iva Bojic, Juan~Murillo Arias, Stanislav
  Sobolevsky, and Carlo Ratti.
\newblock Socioeconomic characterization of regions through the lens of
  individual financial transactions.
\newblock {\em PloS one}, 12(11):e0187031, 2017.

\bibitem{santi2014quantifying}
Paolo Santi, Giovanni Resta, Michael Szell, Stanislav Sobolevsky, Steven~H
  Strogatz, and Carlo Ratti.
\newblock Quantifying the benefits of vehicle pooling with shareability
  networks.
\newblock {\em Proceedings of the National Academy of Sciences},
  111(37):13290--13294, 2014.

\bibitem{hawelka2014geo}
Bartosz Hawelka, Izabela Sitko, Euro Beinat, Stanislav Sobolevsky, Pavlos
  Kazakopoulos, and Carlo Ratti.
\newblock Geo-located twitter as proxy for global mobility patterns.
\newblock {\em Cartography and Geographic Information Science}, 41(3):260--271,
  2014.

\bibitem{belyi2017global}
Alexander Belyi, Iva Bojic, Stanislav Sobolevsky, Izabela Sitko, Bartosz
  Hawelka, Lada Rudikova, Alexander Kurbatski, and Carlo Ratti.
\newblock Global multi-layer network of human mobility.
\newblock {\em International Journal of Geographical Information Science},
  31(7):1381--1402, 2017.

\bibitem{paldino2015urban}
Silvia Paldino, Iva Bojic, Stanislav Sobolevsky, Carlo Ratti, and Marta~C
  Gonz{\'a}lez.
\newblock Urban magnetism through the lens of geo-tagged photography.
\newblock {\em EPJ Data Science}, 4(1):5, 2015.

\bibitem{wang2017structure}
Lingjing Wang, Cheng Qian, Philipp Kats, Constantine Kontokosta, and Stanislav
  Sobolevsky.
\newblock Structure of 311 service requests as a signature of urban location.
\newblock {\em PloS one}, 12(10):e0186314, 2017.

\bibitem{yoshimura2014analysis}
Yuji Yoshimura, Stanislav Sobolevsky, Carlo Ratti, Fabien Girardin, Juan~Pablo
  Carrascal, Josep Blat, and Roberta Sinatra.
\newblock An analysis of visitors' behavior in the louvre museum: A study using
  bluetooth data.
\newblock {\em Environment and Planning B: Planning and Design},
  41(6):1113--1131, 2014.

\bibitem{zhu2018digital}
Enwei Zhu, Maham Khan, Philipp Kats, Shreya~Santosh Bamne, and Stanislav
  Sobolevsky.
\newblock Digital urban sensing: A multi-layered approach.
\newblock {\em arXiv preprint arXiv:1809.01280}, 2018.

\bibitem{reynolds2015gaussian}
Douglas Reynolds.
\newblock Gaussian mixture models.
\newblock {\em Encyclopedia of biometrics}, pages 827--832, 2015.

\bibitem{fortunato2010community}
Santo Fortunato.
\newblock Community detection in graphs.
\newblock {\em Physics reports}, 486(3-5):75--174, 2010.

\bibitem{macqueen1967some}
James MacQueen et~al.
\newblock Some methods for classification and analysis of multivariate
  observations.
\newblock In {\em Proceedings of the fifth Berkeley symposium on mathematical
  statistics and probability}, volume~1, pages 281--297. Oakland, CA, USA,
  1967.

\bibitem{ester1996density}
M~Ester, HP~Kriegel, J~Sander, and Xu~Xiaowei.
\newblock A density-based algorithm for discovering clusters in large spatial
  databases with noise.
\newblock Technical report, AAAI Press, Menlo Park, CA (United States), 1996.

\bibitem{sobolevsky2014general}
Stanislav Sobolevsky, Riccardo Campari, Alexander Belyi, and Carlo Ratti.
\newblock General optimization technique for high-quality community detection
  in complex networks.
\newblock {\em Physical Review E}, 90(1):012811, 2014.

\bibitem{gupta2012integrating}
Manish Gupta, Jing Gao, Yizhou Sun, and Jiawei Han.
\newblock Integrating community matching and outlier detection for mining
  evolutionary community outliers.
\newblock In {\em Proceedings of the 18th ACM SIGKDD international conference
  on Knowledge discovery and data mining}, pages 859--867. ACM, 2012.

\bibitem{duan2009community}
Dongsheng Duan, Yuhua Li, Yanan Jin, and Zhengding Lu.
\newblock Community mining on dynamic weighted directed graphs.
\newblock In {\em Proceedings of the 1st ACM international workshop on Complex
  networks meet information \& knowledge management}, pages 11--18. ACM, 2009.

\bibitem{jolliffe2011principal}
Ian Jolliffe.
\newblock {\em Principal component analysis}.
\newblock Springer, 2011.

\bibitem{golub1971singular}
Gene~H Golub and Christian Reinsch.
\newblock Singular value decomposition and least squares solutions.
\newblock In {\em Linear Algebra}, pages 134--151. Springer, 1971.

\bibitem{miller2012scalable}
Benjamin~A Miller, Nicholas Arcolano, Michelle~S Beard, Jeremy Kepner,
  Matthew~C Schmidt, Nadya~T Bliss, and Patrick~J Wolfe.
\newblock A scalable signal processing architecture for massive graph analysis.
\newblock In {\em 2012 IEEE International Conference on Acoustics, Speech and
  Signal Processing (ICASSP)}, pages 5329--5332. IEEE, 2012.

\bibitem{lakhina2004diagnosing}
Anukool Lakhina, Mark Crovella, and Christophe Diot.
\newblock Diagnosing network-wide traffic anomalies.
\newblock In {\em ACM SIGCOMM computer communication review}, volume~34, pages
  219--230. ACM, 2004.

\bibitem{hinton2006reducing}
Geoffrey~E Hinton and Ruslan~R Salakhutdinov.
\newblock Reducing the dimensionality of data with neural networks.
\newblock {\em science}, 313(5786):504--507, 2006.

\bibitem{wiki:protest}
{Wikipedia contributors}.
\newblock List of protests in the united states by size --- {Wikipedia}{,} the
  free encyclopedia, 2019.
\newblock [Online; accessed 22-October-2019].

\bibitem{bouveyron2007high}
Charles Bouveyron, St{\'e}phane Girard, and Cordelia Schmid.
\newblock High-dimensional data clustering.
\newblock {\em Computational Statistics \& Data Analysis}, 52(1):502--519,
  2007.

\bibitem{abdi2010principal}
Herv{\'e} Abdi and Lynne~J Williams.
\newblock Principal component analysis.
\newblock {\em Wiley interdisciplinary reviews: computational statistics},
  2(4):433--459, 2010.

\bibitem{scikit-learn}
F.~Pedregosa, G.~Varoquaux, A.~Gramfort, V.~Michel, B.~Thirion, O.~Grisel,
  M.~Blondel, P.~Prettenhofer, R.~Weiss, V.~Dubourg, J.~Vanderplas, A.~Passos,
  D.~Cournapeau, M.~Brucher, M.~Perrot, and E.~Duchesnay.
\newblock Scikit-learn: Machine learning in {P}ython.
\newblock {\em Journal of Machine Learning Research}, 12:2825--2830, 2011.

\end{thebibliography}

\end{document}